\documentclass[11pt,letterpaper]{article}
\pdfoutput=1
\usepackage{jcappub}
\usepackage{bbm}
\usepackage{mathrsfs}
\usepackage{slashed}
\usepackage{caption}
\usepackage{epstopdf}
\usepackage[normalem]{ulem}
\usepackage[bottom]{footmisc}
\usepackage{subcaption}
\usepackage{bbold}
\usepackage{titlesec}
\usepackage{threeparttable}
\usepackage{booktabs}
\usepackage{changepage}
\usepackage[utf8]{inputenc}
\usepackage{dsfont} 
\usepackage{grffile}
\usepackage{graphicx}  
\usepackage{dcolumn}   
\usepackage{bm}        
\usepackage{amssymb}   
\usepackage{setspace}
\usepackage{amsmath, amssymb, setspace}
\usepackage{array}
\usepackage{booktabs}
\usepackage{caption}
\usepackage{float}
\usepackage{lmodern}
\usepackage{multirow}
\usepackage{soul}
\usepackage[normalem]{ulem}
\usepackage{braket}
\usepackage{comment}
\usepackage[draft]{pgf}
\usepackage{adjustbox} 
\usepackage{xspace} 
\usepackage{url}
\usepackage{xcolor}
\usepackage{comment}

\usepackage{orcidlink}

\usepackage{indentfirst}
\newcommand{\be}{\begin{equation}}
\newcommand{\ee}{\end{equation}}
\def \d {{\rm d}}

\newcommand{\bea}{\begin{eqnarray}}
\newcommand{\eea}{\end{eqnarray}}

\title{Black Holes as Fermion Factories}

\author{Yifan Chen\,\orcidlink{0000-0002-2507-8272}$^{a}$, Xiao Xue\,\orcidlink{0000-0002-0740-1283}$^{b,c,d}$, and Vitor Cardoso\,\orcidlink{0000-0003-0553-0433}$^{a,e}$}

\affiliation{
$^a$Niels Bohr International Academy, Niels Bohr Institute, Blegdamsvej 17, 2100 Copenhagen, Denmark\\
$^b$Institut de Física d’Altes Energies (IFAE), The Barcelona Institute of Science and Technology, Campus UAB, 08193 Bellaterra (Barcelona), Spain\\
$^c$II. Institute of Theoretical Physics, Universit\"at Hamburg, 22761 Hamburg, Germany\\
$^d$Deutsches Elektronen-Synchrotron DESY, Notkestr. 85, 22607, Hamburg, Germany\\
$^e$CENTRA, Departamento de F\'{\i}sica, Instituto Superior T\'ecnico -- IST, Universidade de Lisboa -- UL, Avenida Rovisco Pais 1, 1049 Lisboa, Portugal}
\emailAdd{yifan.chen@nbi.ku.dk}
\emailAdd{xxue@ifae.es}
\emailAdd{vitor.cardoso@nbi.ku.dk}

\abstract{Ultralight bosons near rotating black holes can undergo significant growth through superradiant energy extraction, potentially reaching field values close to the Planck scale and transforming black holes into effective transducers for these fields. The interaction between boson fields and fermions may lead to parametric production or Schwinger pair production of fermions, with efficiencies significantly exceeding those of perturbative decay processes. Additionally, the spatial gradients of scalar clouds and the electric components of vector clouds can accelerate fermions, resulting in observable fluxes. This study considers both Standard Model neutrinos and dark sector fermions, which could contribute to boosted dark matter. Energy loss due to fermion emissions can potentially quench the exponential growth of the cloud, leading to a saturated state. This dynamic provides a framework for establishing limits on boson-neutrino interactions, previously constrained by neutrino self-interaction considerations. In the saturation phase, boson clouds have the capacity to accelerate fermions to TeV energies, producing fluxes that surpass those from atmospheric neutrinos near black holes. These fluxes open new avenues for observations through high-energy neutrino detectors like IceCube, as well as through dark matter direct detection efforts focused on targeted black holes.}


\arxivnumber{DESY-23-109}

\begin{document}
\maketitle
\flushbottom

\section{Introduction}

Hypothetical bosons with ultralight masses are promising candidates in the search for physics beyond the Standard Model. Their potential to explain phenomena such as the smallness of the neutron electric dipole moment~\cite{Preskill:1982cy, Abbott:1982af, Dine:1982ah} and their emergence from fundamental theories that incorporate extra dimensions~\cite{Svrcek:2006yi, Abel:2008ai, Arvanitaki:2009fg, Goodsell:2009xc} make them particularly intriguing. When these bosons form dark matter and possess masses below $\mathcal{O}(1)\,\text{eV}$, they manifest as coherent waves due to their typically high occupation numbers~\cite{Hu:2000ke}. The interactions between an ultralight boson background and Standard Model particles can generate observable signals, distinguished by their frequency and proportional to the energy density or field value of the bosons.

When the Compton wavelength of these bosons is comparable to the gravitational radius of a rapidly rotating black hole (BH), a bound state may form through the superradiance mechanism, which extracts rotational energy from the BH~\cite{Penrose:1971uk,ZS,Detweiler:1980uk,Cardoso:2005vk,Dolan:2007mj,Brito:2015oca}. This leads to the condensation of a bosonic structure, known as a boson cloud, in the exterior of the BH~\cite{Detweiler:1980uk,Cardoso:2005vk,Dolan:2007mj}. These superradiant clouds can amass up to approximately $10 \%$ of the BH mass~\cite{Brito:2014wla,East:2017ovw,Herdeiro:2021znw}, reaching field values near the grand unification theory (GUT) scale~\cite{Chen:2022kzv}, significantly surpassing the local dark matter field value. Thus, Kerr BHs can serve as potent transducers for ultralight bosons~\cite{Brito:2015oca}. The detection of superradiant phenomena does not require that ultralight bosons constitute the majority of dark matter and can be achieved through various methods, including BH spin-down~\cite{Arvanitaki:2010sy,Arvanitaki:2014wva,Brito:2014wla,Baryakhtar:2017ngi,Brito:2017zvb,Cardoso:2018tly,Davoudiasl:2019nlo,Brito:2020lup,Stott:2020gjj,Unal:2020jiy,Saha:2022hcd}, gravitational wave signals from boson clouds~\cite{Arvanitaki:2010sy,Yoshino:2012kn,Yoshino:2013ofa,Arvanitaki:2014wva,Yoshino:2015nsa,Baryakhtar:2017ngi,Brito:2017wnc,Brito:2017zvb,Isi:2018pzk,Siemonsen:2019ebd,Sun:2019mqb,Palomba:2019vxe,Brito:2020lup,Zhu:2020tht,Tsukada:2020lgt,Yuan:2021ebu,KAGRA:2021tse,Yuan:2022bem,Chen:2022kzv,Brito:2023pyl}, and axion cloud-induced birefringence~\cite{Chen:2019fsq,Yuan:2020xui,Chen:2021lvo,Chen:2022oad}.

For a boson field with minimal gravitational interaction, the exponential superradiant growth persists until the BH spin reaches the threshold for spin-down, with the cloud mass growing to approximately $10\%$ of the BH mass. Afterward, the cloud slowly decays via gravitational wave emission. As a result, BH spin measurements and gravitational wave detections from the cloud are the two primary observables for probing ultralight bosons with masses satisfying the superradiance condition.

On the other hand, boson self-interactions and interactions with other fields can potentially transform the bound state of the cloud into fluxes escaping to infinity, thereby quenching the exponential growth. When the energy leakage rate becomes comparable to the superradiant energy gain rate from the BH, the cloud enters a quasi-equilibrium state. Previous studies have demonstrated that both axion self-interactions~\cite{Arvanitaki:2010sy,Yoshino:2012kn,Gruzinov:2016hcq,Fukuda:2019ewf,Baryakhtar:2020gao,Omiya:2020vji,Omiya:2022mwv,Omiya:2022gwu,Collaviti:2024mvh,Takahashi:2024fyq,Aurrekoetxea:2024cqd} and axion-photon couplings~\cite{Rosa:2017ury,Boskovic:2018lkj,Ikeda:2018nhb,Spieksma:2023vwl} can lead to a saturation phase following the exponential growth stage.

Studying the potential interactions involving ultralight bosons is therefore crucial for understanding the superradiance mechanism. These interactions can terminate exponential growth before significant backreaction on the BH spin occurs, potentially making BH spin-down negligible and gravitational wave emission inefficient. Detailed calculations of energy emission rates and the saturation cloud mass are required to determine the exclusion region for interaction strengths, which typically exclude sufficiently weak couplings. Furthermore, the total cloud mass during saturation, a macroscopic quantity, is directly related to the microscopic interactions responsible for energy emission.

The particle production from a superradiant cloud can be significantly enhanced due to the GUT-scale field values near BHs, providing a unique testing ground for non-perturbative physics. Such particle production is traditionally studied in the context of early universe cosmology, where a dense, coherently oscillating scalar background induces parametric resonance, leading to enhanced matter production~\cite{Greene:1998nh, Greene:2000ew}. A key difference between the superradiant cloud and the cosmological scalar background lies in the non-uniform spatial distribution of the cloud’s wavefunction. Additionally, in the realm of strong-field quantum electrodynamics, a strong electromagnetic field background is known to produce matter-antimatter pairs via the Schwinger pair production mechanism~\cite{Schwinger:1951nm}.

In this study, we demonstrate that both scalar and vector clouds surrounding BHs, by leveraging their significant field strengths, can produce substantial fermion fluxes through parametric excitation~\cite{Greene:1998nh, Greene:2000ew} and Schwinger pair production~\cite{Schwinger:1951nm}, respectively. Furthermore, the non-uniform scalar cloud background and the electric components of the vector cloud can further accelerate the produced fermions, resulting in high-energy fluxes. A dark-sector fermion flux with velocities exceeding those of cold dark matter is commonly referred to as boosted dark matter, which can be probed by terrestrial dark matter and neutrino detectors~\cite{DEramo:2010keq,Huang:2013xfa,Agashe:2014yua,Kopp:2015bfa,Bhattacharya:2016tma,Kamada:2017gfc,Kachulis:2017nci,Chatterjee:2018mej,Kamada:2018hte,McKeen:2018pbb,Arguelles:2019xgp,Kamada:2019wjo,Berger:2019ttc,Abi:2020evt}.

We further consider the interaction between ultralight bosons and neutrinos. Among the various interactions between ultralight bosons and the Standard Model, the interaction with neutrinos is particularly well-motivated, primarily due to its implications for neutrino mass generation~\cite{Gelmini:1980re} and grand unification theories~\cite{Georgi:1974sy, Pati:1974yy, Mohapatra:1974hk}. However, exploring these interactions poses significant challenges due to the difficulties in neutrino production and detection, with neutrino self-interaction being the primary observable~\cite{Berryman:2022hds}. Despite these challenges, when bosons meet the superradiance condition, their resulting neutrino fluxes from nearby BHs have the potential to surpass the ambient diffusive neutrino background~\cite{Vitagliano:2019yzm}, opening new possibilities for high-energy neutrino detection.

The structure of this paper is as follows: Section~\ref{sec:FPSC} details how scalar or vector clouds with significant field values can produce fermions more efficiently than perturbative decay and accelerate them to higher energies. This section also explores the possibility of reaching a saturation state where total fermion emissions balance the superradiant energy input from BHs. Section~\ref{sec:SMNBI} examines how measurements of BH spins can exclude parameter spaces with weak coupling strengths. Section~\ref{sec:NBDMF} considers potential observational channels for detecting fermion fluxes in the viable parameter space. Section~\ref{sec:Discussion} summarizes the results and discusses prospects for future observations.

\section{Fermion Production from Superradiant Clouds}\label{sec:FPSC}

This section begins with a brief introduction to superradiance, covering the superradiance conditions for both scalar and vector fields, the wavefunction of the superradiant ground state, and possible mechanisms to quench exponential growth. We then focus on matter production from boson condensates, including parametric production and Schwinger pair production of fermions. Next, we calculate and simulate the acceleration trajectories of these particles within the cloud, considering the non-uniformity of scalar clouds and the electric components of vector clouds. Finally, we discuss the potential saturation of the cloud state, resulting from the balance between the BH’s superradiant energy input and the emission of fermions.

\subsection{Superradiance}\label{sec:SR}
Ultralight bosons, often predicted in theories incorporating extra dimensions~\cite{Svrcek:2006yi, Abel:2008ai, Arvanitaki:2009fg, Goodsell:2009xc}, are promising candidates for cold dark matter~\cite{Preskill:1982cy, Abbott:1982af, Dine:1982ah, Nelson:2011sf}, notable for their wave-like properties~\cite{Hu:2000ke}. These properties facilitate the formation of bound states known as gravitational atoms around BHs, resulting from gravitational attraction~\cite{Detweiler:1980uk}.
These states are characterized by unique quantum numbers and are influenced by the gravitational fine-structure constant, denoted as $\alpha \equiv G_{\rm N} M_{\rm BH} \mu$~\cite{Detweiler:1980uk, Brito:2015oca}, where $G_{\rm N}$ is Newton's constant, $M_{\rm BH}$ is the BH mass, and $\mu$ is the boson mass.

Around Kerr BHs, gravitational atoms with specific quantum numbers that corotate with the BH can undergo exponential growth by extracting energy through BH rotation, a phenomenon known as superradiance~\cite{Penrose:1971uk, ZS, Detweiler:1980uk, Cardoso:2005vk, Dolan:2007mj, Brito:2015oca}. The parameter $\alpha$ must satisfy the superradiant condition formulated as:
\be \frac{\alpha}{m}<\frac{a_J}{2\left(1+\sqrt{1-a_J^2}\right)},\ee
where $a_J$ represents the BH's dimensionless spin parameter, and $m$ is the azimuthal quantum number of the gravitational atom, with $m>0$ indicating corotation with the BH.

Since ground states in superradiance exhibit the fastest growth rates, our analysis focuses on these states. Under the Newtonian approximation ($\alpha \ll 1$), the wavefunctions for both scalar and vector fields are expressed as follows:~\cite{Detweiler:1980uk,Brito:2015oca,Baryakhtar:2017ngi,Siemonsen:2022ivj}
\be
\begin{split} \phi &= \Psi_0(t)\, e^{1-\alpha^2 r/(2 r_g)}\,\frac{\alpha^2 r}{2 r_g} \cos \left(\mu t - \varphi\right) \sin \theta,\\  
{A}^{\prime\,\mu} &= \Psi_{0}(t)\, e^{-\alpha^2 r/r_g}\, \left( \alpha \sin \theta \sin (\mu t - \varphi) , \cos \left(\mu t\right), \sin \left(\mu t\right) , 0 \right),    \end{split}\label{eq:GAWF} 
\ee
utilizing Boyer-Lindquist coordinates $(t,r,\theta,\varphi)$. Here, $\phi$ and ${A}^{\prime\,\mu}$ represent the scalar field and the components of the vector field in the unitary gauge, respectively. $\Psi_0$ denotes the peak field value within the bosonic cloud, and $r_g \equiv G_{\rm N} M_{\rm BH}$ represents the gravitational radius. The Newtonian approximation for these wavefunctions is valid in the region where $r \gg r_g$. Since the cloud size, or the Bohr radius, is on the order of $r_g / \alpha^2$, this approximation is appropriate in the regime of small $\alpha$, where the majority of the cloud resides at distances satisfying $r \gg r_g$. In this study, we consider $\alpha < 0.3$ to ensure that the condition for the Newtonian approximation is met.

The superradiant growth rate, $\Gamma_{\rm SR}$, is approximated as $\alpha^8 a_J \mu/24$ for scalar fields and $4\alpha^6 a_J \mu$ for vector fields when $\alpha \ll 1$~\cite{Detweiler:1980uk}. Notably, since the exponential growth is insensitive to the initial conditions, the detection of superradiant phenomena does not require that ultralight bosons constitute the majority of dark matter.

In cases where a boson minimally interacts with gravity, energy extraction from the BH rotation continues until the mass of the boson cloud, denoted as $M_{\rm cloud}$, approximates about $10\%$ of the BH mass~\cite{Brito:2014wla,East:2017ovw,Herdeiro:2021znw}. Without an external influx of angular momentum, reaching this $10\%M_{\rm BH}$ threshold causes the BH’s spin to decrease to a point where superradiance is no longer supported. Integrating the energy density of the cloud over its entire volume allows us to estimate the total mass of the cloud, $M_{\rm cloud} \approx 186 \Psi^{2}_{0}/(\alpha^3 \mu)$ for scalar fields and $\pi \Psi^{2}_{0}/(\alpha^3 \mu)$ for vector fields~\cite{Chen:2022kzv}. When $M_{\rm cloud}$ constitutes $10\%$ of the BH mass, $M_{\rm BH}$, the field values approach Planck scales:
\be \Psi_0= \left(\frac{\alpha}{0.2}\right)^2 \left(\frac{M_{\rm cloud}}{10\% M_{\rm BH}}\right)^{1/2} \times \left\{\begin{array}{ll}
 1.1\times10^{16}\, {\rm GeV},\qquad \text{Scalar} \\
8.7\times10^{16}\, {\rm GeV}.\qquad \text{Vector}
\end{array}\right. . \label{eq:MC10}
\ee
Hereafter, we define $\Psi_0^{10\%}$ as the maximum permissible field value by setting $M_{\rm cloud}/M_{\rm BH}=0.1$.

Exponential growth of superradiant clouds can be halted before causing significant backreaction on the BH if there is an interaction between the boson cloud and other fields. In such scenarios, the cloud reaches a quasi-equilibrium state where these interactions induce a continuous outflow of energy. This outflow balances the energy gain from superradiance, maintaining an equilibrium state. Examples of such interactions include quartic self-interactions~\cite{Arvanitaki:2010sy,Yoshino:2012kn,Gruzinov:2016hcq,Fukuda:2019ewf,Baryakhtar:2020gao,Omiya:2020vji,Omiya:2022mwv,Omiya:2022gwu,Collaviti:2024mvh,Takahashi:2024fyq,Aurrekoetxea:2024cqd}, axion-photon coupling~\cite{Rosa:2017ury,Boskovic:2018lkj,Ikeda:2018nhb,Spieksma:2023vwl}, and electron-positron pair production~\cite{Fukuda:2019ewf,Blas:2020kaa,Siemonsen:2022ivj}. Notably, if the coupling strengths are sufficiently weak, the BH may spin down before the cloud reaches saturation, allowing high-spin BHs to exclude these regions of parameter space.

One notable example involves the axion, characterized by a cosine potential $V \approx -\mu^2 f_\phi^2\\ \cos(\phi/f_\phi)$, where $f_\phi$ denotes the axion decay constant. The self-interactions stemming from this potential, particularly the quartic term $V \supset - \lambda \phi^4 \equiv -\phi^4 \mu^2/(24 f_\phi^2)$, quench the exponential growth into a saturated state. In this state, the superradiant ground mode field value stabilizes at $\Psi_0^\lambda \approx \alpha f_\phi/2$, resulting in steady axion waves radiating towards infinity~\cite{Yoshino:2012kn,Gruzinov:2016hcq,Baryakhtar:2020gao}. This mechanism sets another threshold for the total cloud mass, distinct from the spin-down requirement of $10\% M_{\rm BH}$.

This study investigates the direct couplings between a bosonic cloud and fermions. The intense field within the cloud can trigger the parametric production of fermions, which may then undergo further acceleration as they propagate through the cloud. This process potentially leads to a saturated cloud state that efficiently converts the rotational energy of the BH into fermion emissions.

\subsection{Particle Production from Boson Condensates}\label{sec:PRBC}

In this section, we review parametric matter production from oscillating scalar backgrounds and Schwinger pair production from vector backgrounds. We also discuss how these mechanisms apply to superradiant clouds.

\subsubsection{Parametric Matter Production from Scalar Backgrounds}
We begin by considering a bosonic field $\chi$ in a time-varying background~\cite{Zeldovich:1971mw,Kofman:1997yn,Greene:1998nh,Greene:2000ew}. The equation of motion for $\chi$ in the frequency domain is given by:
\be \ddot{\chi}_k+\omega_k(t)^2 \chi_k=0. \label{eq:Xk}\ee
Here, $\omega_k(t)$ represents the time-varying energy, modulated by the background fields to which $\chi$ couples. The solutions to this equation in the adiabatic representation are expressed as~\cite{Kofman:1997yn}:
\be
\chi_k(t)=\frac{\alpha_k(t)}{\sqrt{2 \omega_k}} e^{-i \int^t_0 \omega_k\d t}+\frac{\beta_k(t)}{\sqrt{2 \omega_k}} e^{+i \int^t_0 \omega_k\d t}.\ee
The coefficients $\alpha_k$ and $\beta_k$ evolve according to:
\be
\dot{\alpha}_k=\frac{\dot{\omega}_k}{2 \omega_k} e^{+2 i \int^t_0 \omega_k\d t} \beta_k, \quad \dot{\beta}_k=\frac{\dot{\omega}_k}{2 \omega_k} e^{-2 i \int^t_0 \omega_k\d t} \alpha_k, \label{eq:alphabeta}
\ee
and they adhere to the normalization condition $|\alpha_k|^2 - |\beta_k|^2 = 1$. At $t=0$, the vacuum state is defined by $\alpha_k = 1$ and $\beta_k = 0$. The quantities $\alpha_k(t)$ and $\beta_k(t)$ correspond to the coefficients of the Bogoliubov transformation, which diagonalize the Hamiltonian at each moment $t$. The particle occupation number for a momentum mode $k$ is given by $n_k = |\beta_k|^2$, and the particle number density per unit volume is:
\be 
n_\chi=\frac{1}{2 \pi^2} \int_0^{\infty} d k k^2\left|\beta_k\right|^2 .
\ee
For $|\beta_k| \ll 1$, the solution for Eq.~(\ref{eq:alphabeta}) can be approximated as:
\be
\beta_k \simeq  \int_0^t\d t^{\prime} \frac{\dot{\omega}}{2\omega} \exp \left(-2 i \int_0^{t^{\prime}}\d t^{\prime \prime} \omega\left(t^{\prime \prime}\right)\right).
\ee
Consequently, efficient particle production occurs when the non-adiabatic condition is satisfied:
\be \left|\frac{\dot{\omega}}{\omega^2}\right| \gg 1.\label{eq:nac}\ee

Using the interaction term $g_{S\chi}^2\phi^2\chi^2/2$ as an example, let us consider $\phi = \phi_0 \cos(\mu t)$ to represent the coherently oscillating background scalar field with amplitude $\phi_0$ and frequency $\mu$, and $g_{S\chi}$ to denote the coupling constant. The frequency relation for $\chi$ is~\cite{Kofman:1997yn}
\be \omega_\chi^2 = k^2 + m_\chi^2 + g_{S\chi}^2 \phi_0^2 \cos^2(\mu t).\label{eq:ochi}\ee
Here, $m_\chi$ represents the bare mass term of $\chi$. We concentrate on the regime where $g_{S\chi}^2 \phi_0^2/\mu^2\gg1$,  and the mass $m_\chi$ can be considered negligible. Within a narrow region of $\phi$ during a single oscillation period, momentum values below the typical scale $k_\star = \sqrt{g_{S\chi} \phi_0 \mu}/2$ satisfy the non-adiabatic condition (\ref{eq:nac}). Consequently, the occupation number $n_k$ for $k < k_\star$ is exponentially produced with a time dependence of approximately $\exp(2 \mu \tau_k t)$, where $\tau_k \approx \mathcal{O}(0.1)$~\cite{Kofman:1997yn}.

A fermionic field $\psi$ with a Yukawa-like coupling $g_{S} \phi \bar{\psi} \psi$ and a bare mass $m_\psi$ follows a frequency relation for their mode function described by~\cite{Greene:1998nh,Greene:2000ew}:
\be \omega_\psi^2 = k^2 + \left( m_\psi + g_{S} \phi_0 \cos(\mu t) \right)^2 - i g_S \mu \phi_0 \sin (\mu t).\label{eq:opsi}\ee
In regions with a dense boson background, where the resonance parameter $q \equiv g_{S}^2 \phi_0^2/\mu^2$ is significantly high and $g_{S} \phi_0 \gg m_\psi$, the non-adiabatic condition is met when the effective mass term, $m_{\rm eff} \equiv m_\psi + g_{S} \phi_0 \cos(\mu t)$, crosses zero. During each crossing, the background substantially contributes to the $k$-mode with a factor of $|\beta_k|^2 = e^{-\pi k^2/(g_{S} \phi_0 \mu)} + \cdots$, where $\cdots$ accounts for contributions from previously produced fermions~\cite{Greene:1998nh,Greene:2000ew}. Exponential growth is precluded due to Pauli blocking. Instead, a fermion sphere with a radius approximately $k_F = \sqrt{g_{S} \phi_0 \mu}/2$ forms immediately after each interaction.

For fermions produced from a locally dense scalar condensate, those generated during the previous interaction are accelerated to much higher energy scales, a topic further discussed in the subsequent subsection, without hindering new production. Consequently, the average pair production rate per unit volume is estimated as:
\be \Gamma_{\phi\psi} \approx \frac{1}{2\pi^2} \frac{k_{F}^3}{3} \frac{\mu}{\pi} = \frac{g_{S}^2\phi_0^2 \mu^2}{48\pi^3} \sqrt{\frac{\mu}{g_{S}\phi_0}}. \ee
The factor $\pi/\mu$ represents the time interval between two zero crossings. Additionally, when the de Broglie wavelength of the initially produced fermions, approximately $2\pi/k_\star$, is considerably shorter than the condensate size, the finite size of the boson background can be disregarded in calculating the production rate, akin to the production of millicharged particles from superconducting cavities~\cite{Berlin:2020pey}.

This parametric excitation of fermion and anti-fermion pairs is more efficient than perturbative decay, especially since an ultralight scalar cannot decay into fermions heavier than itself through perturbative processes.

One fundamental distinction between boson and fermion production lies in the requirement for the bare mass term. In Eq.~(\ref{eq:opsi}), the effective mass is the linear sum of the bare mass and the oscillating term, necessitating only that $g_{S} \phi_0 \gg m_\psi$ to facilitate parametric excitation~\cite{Greene:2000ew}. Conversely, for bosons, the bare mass $m_\chi$ must be smaller than $k_\star$ due to their analogous roles in influencing the dynamics as detailed in Eq.~(\ref{eq:ochi}). Consequently, heavier fermions can be parametrically excited compared to bosons.

\subsubsection{Schwinger Pair Production from Vector Backgrounds}

We next consider a charged scalar field $\sigma$ in a static electric background along the $z$-axis, specifically $A_z = E t$ in the temporal gauge. The dispersion relation is given by
\be \omega_\sigma^2 = \left(k_z - e E t \right)^2 +k_\perp^2 + m_\sigma^2,\ee
where $e$ represents the charge of $\sigma$, $k_\perp$ denotes its momentum transverse to the $z$-axis, and $m_\sigma$ is its mass. When $eE \gg m_\sigma^2$, we can disregard the mass term and anticipate the production of relativistic $\sigma$ particles. Considering the non-adiabatic condition (\ref{eq:nac}) once again, we observe that modes with $0 < k_z \leq \sqrt{eE}$ and $k_\perp \ll \sqrt{eE}$ satisfy it at $t = 0$. These modes continue to satisfy the condition until approximately $t \approx 1/\sqrt{eE}$, at which point the corresponding occupation number becomes saturated due to the balance between production and acceleration towards higher energies. The presence of $k_\perp^2 + m_\sigma^2$ introduces an exponential suppression factor to the occupation number, providing an approximation to the production rate~\cite{grib1994vacuum}:
\be \begin{split}
\Gamma_{Schw} \approx& \ \sqrt{eE} \int \frac{\d^3\vec{k}}{(2\pi)^3} \exp\left( - \frac{\pi (m_\sigma^2 + k_\perp^2)}{eE} \right) \text{H}\left(\sqrt{eE}-k_z \right)  \text{H}\left( k_z \right) \\
=& \ \frac{e^2 E^2}{8\pi^3} \exp\left(-\frac{\pi m_\sigma^2}{eE}\right),
\end{split}\ee
where $\text{H}$ is the Heaviside function. The factor $1/\sqrt{eE}$ denotes the production timescale. A more complete sum introduces an additional factor of $\Sigma_{n=1}^{\infty} 1/n^2 = \pi^2/6$ when $eE \gg m_\sigma^2$~\cite{Schwinger:1951nm}.

For a localized vector condensate characterized by a mass $\mu$, the Schwinger pair production rate is independent of whether the produced particles are bosons or fermions, as the occupation number remains low for $k \leq k_\star = \sqrt{eE}$. Any produced pair is immediately accelerated to higher momenta. Additionally, the de Broglie wavelength, approximately $2\pi/k_\star$, and the production timescale are notably shorter than the condensate’s size and its oscillation period, given that $k_\star \gg m_\psi \gg \mu$. In this context, the finite size and periodic oscillation of the condensate can be disregarded when considering the production rate, which simplifies the analysis of the condensate’s particle dynamics.

\subsubsection{Fermion Production from Boson Clouds}

We now turn our attention to the superradiant scalar and vector clouds discussed in Sec.~\ref{sec:SR}, focusing on their interactions with fermions $\psi$ through Yukawa-like and Abelian gauge interactions, respectively. Within both types of clouds, fermion pairs can spontaneously emerge, driven by the light masses of the fermions and the substantial field values of the bosons.

The spatially dependent scalar field $\phi$ introduces a coherently oscillating mass term, $g_{S} \phi$, to the fermions, in addition to their bare mass, $m_\psi$. When the product $g_{S} \phi_0$ surpasses $m_\psi$, the effective mass, $m_{\rm eff} \equiv g_{S} \phi + m_\psi$, oscillates through zero twice each period. At each zero-crossing, fermions undergo parametric excitation, populating a fermion sphere with a radius $k_F \approx \sqrt{g_{S} \phi_0 \mu}/2$. This calculation assumes that the fermion’s de Broglie wavelength, approximately $1/k_F$, is significantly shorter than the cloud’s size, approximately $1/(\alpha\mu)$, allowing the neglect of finite-size effects. The scalar cloud also exhibits an oscillatory component in its wavefunction, dependent on the azimuthal angle, i.e., $\phi \propto \cos(\mu t - \varphi)$. This dependence stems from the orbital angular momentum of the cloud. Notably, at a specific time $t$, the non-adiabatic condition is triggered on a plane approximately aligned with $\varphi = \mu t \pm \pi/2$. Consequently, the cloud’s oscillation induces a periodic rotation of the production plane around the spin axis of the BH.

Similarly, vector clouds with gauge intereaction $g_V {A}^{\prime\,\mu} \bar{\psi} \gamma_\mu \psi$ may produce fermions through Schwinger pair production when the electric field strength, $E_{A^\prime} \approx \mu |\vec{A}^\prime|$, is high enough that $g_{V} E_{A^\prime} \gg m_\psi^2$~\cite{Schwinger:1951nm}, where $g_V$ denotes the interaction strength. In this scenario, the produced fermions carry momentum approximately $\sqrt{g_V E_{A^\prime}}$, with their directions either aligned or anti-aligned with the electric field direction. These particles are then accelerated by the electric field to energies about $\omega^\psi_{\rm acc} \sim g_V E_{A^\prime}/\mu$, as will be explored further in the next subsection. In this context, the phenomenon of Pauli blocking is negligible, primarily due to the significant difference between $\omega^\psi_{\rm acc}$ and the initial momentum $\sim \sqrt{g_V E_{A^\prime}}$.

In summary, the average pair production rates per unit volume for each fermion state in these processes are given by:
\be \Gamma_{S\psi} \approx \frac{g_{S}^2\phi_0^2 \mu^2}{48\pi^3} \sqrt{\frac{\mu}{g_{S}\phi_0}}, \qquad 
\Gamma_{V\psi} \approx \frac{g_{V}^2 E_{A^\prime}^2}{48\pi},
\label{eq:Ganu}\ee
where $\phi_0$ and $E_{A^\prime}$ are influenced by the cloud wavefunctions, which exhibit spatial and temporal variations.

\subsection{Fermion Acceleration and Resulting Fluxes from Boson Clouds}

To analyze the emitted fermion spectrum and energy extraction from superradiant clouds, it is essential to consider their propagation immediately after production. Both non-uniform scalar clouds and the electric components of vector clouds can exert forces on the produced fermions, accelerating them to significantly higher energies. As discussed in Sec.~\ref{sec:PRBC}, we consider the scenario where the fermion de Broglie wavelength, $< 1/\sqrt{g_{S/V} \Psi_0 \mu}$, is considerably smaller than both the cloud size, $\sim {r_g/\alpha^2 = 1/(\mu\alpha)}$, and the oscillation timescale, $\sim 1/\mu$. This allows us to model each fermion as a point-like particle and use the worldline equation of motion to study their propagation.

Upon their production, fermions begin their propagation through the bosonic field surrounding the BH. This scenario is akin to analyzing geodesics with a variable mass term, $m_{\rm eff} = g_{S}\phi + m_\psi$ for scalar clouds, or considering a charged particle influenced by both the electromagnetic field of the vector cloud and the BH’s gravitational potential. The worldline action governing the fermion’s trajectory is described by:
\be
S_{\psi} = \left\{  \begin{aligned}
&-\int \d \tau\, |m_{\rm eff}|\,\sqrt{-g_{\alpha\beta}u^{\alpha}_\psi u^{\beta}_\psi},\qquad \qquad &\text{Scalar} \\
&-\int \d \tau \left(m_\psi \, \sqrt{-g_{\alpha\beta}u^{\alpha}_\psi u^{\beta}_\psi}\mp g_{V}A^{\prime}_{\beta}u^{\beta}_{\psi}\right),\qquad \qquad &\text{Vector}
\end{aligned}\right. 
,\label{eq:WLA}
\ee
where $g_{\alpha\beta}$ is the Kerr metric and $u^{\alpha}_\psi$ is the fermion's four-velocity. The symbol $\mp$ differentiates fermion from antifermion. From this action, the Euler-Lagrange equation is derived as:
\be
    \frac{\d u_\psi^{\alpha}}{\d \tau} = -\Gamma^{\alpha}_{\kappa\beta}u_\psi^{\kappa}u_\psi^{\beta} 
 +   \left\{  \begin{aligned}
&- (g^{\alpha\beta}+u_\psi^{\alpha}u_\psi^{\beta})\frac{\nabla_{\beta}m_{\rm eff}}{m_{\rm eff}},\qquad \qquad &\text{Scalar} \\
& \pm u^{\beta}_{\psi}\,\frac{g_{V}\,F^{\prime\alpha}_{\,\,\,\,\beta}}{m_{\psi}},\qquad \qquad &\text{Vector}
\end{aligned}\right. ,\label{eq:ELEu}
\ee
where $\tau$ signifies the proper time, the Christoffel symbols, $\Gamma^{\alpha}_{\kappa\beta}$, pertain to the Kerr metric, and $F^{\prime\alpha}_{\,\,\,\,\beta}$ represents the field tensor for ${A}^{\prime\,\mu}$. 
By applying the relations $\d t =u^0_\psi \d \tau$ and $p^{\alpha}_\psi=|m_{\rm eff}|u^{\alpha}_\psi$ (scalar) or $m_\psi u^{\alpha}_\psi$ (vector), we can reformulate this equation as
\be \frac{\d p^\alpha_\psi}{\d t} = -\frac{1}{p^0_\psi}\Gamma^{\alpha}_{\kappa\beta}\, p_\psi^{\kappa}\, p_\psi^{\beta} + \left\{  \begin{aligned}
&- \nabla^{\alpha} m_{\rm eff}^2/(2p_\psi^0),\qquad \qquad &\text{Scalar} \\
&\pm g_V (\vec{E}_{A^\prime} + \vec{v}_\psi\times\vec{B}_{A^\prime} ),\qquad \qquad &\text{Vector}
\end{aligned}\right. 
,\label{eq:wleom}\ee 
where $p_\psi^\alpha$ represents the fermion's four-momentum and $\vec{v}_\psi$ its velocity.
The electric and magnetic fields, $\vec{E}_{A^\prime}$ and $\vec{B}_{A^\prime}$, are defined by $\vec{E}_{A^\prime}\equiv -\nabla A^{\prime 0} - \partial_t \vec{A}^\prime$ and $\vec{B}_{A^\prime}\equiv \nabla \times \vec{A^\prime}$, respectively.

In both scenarios, fermions can be accelerated to energies on the order of $g_{S/V} \Psi_0$, significantly surpassing their initial momentum, approximately $\sqrt{g_{S/V} \Psi_0 \mu}$. Consequently, the Pauli blocking effect can be disregarded for both the production rate and the acceleration process. Furthermore, considering that the production timescale and the initial de Broglie wavelength of the fermion, approximately $1/\sqrt{g_{S/V} \Psi_0 \mu}$, are considerably shorter than the cloud’s oscillation period, roughly $2\pi/\mu$, and the cloud’s size, approximately $1/(\alpha\mu)$, the effects of the cloud’s finite size and temporal variance can be safely overlooked. This is similar to the production mechanisms of millicharged particles from superconducting cavities~\cite{Berlin:2020pey}.

Additionally, we should ensure that the backreaction of fermions on the bosonic cloud can be neglected. Around a vector field, the recoil power on a relativistic fermion in the strong field limit $g_V E_{A^\prime} \gg m_\psi^2$ is typically on the order of $g_{V}^{8/3} E_{A^\prime}^{2/3} \omega_{\rm acc}^{\psi\,2/3} \sim g_{V}^{10/3} \Psi_0^{4/3} \mu^{2/3}$~\cite{Erber:1966vv,1966siiz.book.....S,Blackburn:2019rfv}. Compared to the boson cloud’s acceleration power $\sim g_{V}\Psi_0\mu$, it requires $g_V \ll (\mu/\Psi_0)^{1/7}$. For astrophysical BHs, superradiant $\mu$ is between $10^{-21}$ to $10^{-11}$\,eV and the field value $\Psi_0 < 10^{17}$\,GeV\, and a value of $g_V \ll 10^{-8}$ can easily satisfy this condition.

The forces influencing fermion trajectories in Eq.~(\ref{eq:wleom}) include gravitational deflection and either the scalar force~\cite{Uzan:2020aig} or electromagnetic forces from a vector cloud. For scalar clouds, we analyze their comparative impacts in Cartesian coordinates $(t, x, y, z)$:
\bea
-\frac{\Gamma^{\alpha}_{\kappa\beta}\, p_\psi^{\kappa}\, p_\psi^{\beta}}{m^2_{\rm eff}/r_g} &\approx& -\frac{r_g^2}{r^2}\, \left( \mathcal{O}(1)+\mathcal{O}(\frac{r_g}{r}) \right),\\
-\frac{\vec{\nabla} m^2_{\rm eff}}{m^2_{\rm eff}/r_g} &=& 
 \alpha^2\,\hat{r} - \frac{2\, r_g}{r\cos(\alpha t - \phi)\sin\theta}\, \hat{n}_{\perp} + \cdots.\label{app:terms}
\eea
Here, $\cdots$ denotes the minor impact of the fermion's bare mass term $m_\psi$ in regions where $g_{S} \phi_0 \gg m_\psi$. The vector $\hat{n}_{\perp} \equiv (\cos(\alpha t),\sin(\alpha t),0)$ represents a unit directional vector rotating in the $x-y$ plane. It becomes evident that in most parts of the cloud, the scalar force exceeds gravitational lensing in influence. For vector clouds, a similar quantitative assessment applies in the cloud's dominant region $(r^2>r_g^2/\alpha)$. Consequently, we will neglect the gravitational deflection's effect in our simulation.

We proceed by simulating neutrino trajectories and capturing the flux at infinity, introducing dimensionless quantities for convenience:
{\footnotesize
\be
(\tilde{t}, \tilde{x}, \tilde{y}, \tilde{z}) \equiv \left(\frac{t}{r_g}, \frac{x}{r_g},\frac{y}{r_g},\frac{z}{r_g}\right), \  \tilde{p}^{\alpha}_{\psi} \equiv \frac{p^{\alpha}_{\psi}}{g_{S/V}\Psi_0},\ \tilde{m}_{\psi} \equiv \frac{m_{\psi}}{g_{S/V}\Psi_0}, \ 
    \tilde{m}_{\rm eff} \equiv \frac{m_{\rm eff}}{g_S\Psi_0}, \ \vec{\tilde{E}}_{A^\prime} \equiv \frac{\vec{E}_{A^\prime}}{\Psi_0}, \ \vec{\tilde{B}}_{A^\prime} \equiv \frac{\vec{B}_{A^\prime}}{\Psi_0}.\ee}
This redefinition allows us to describe the trajectories as follows:
\be 
\frac{\d \vec{\tilde{x}}}{\d \tilde{t}}= \frac{\vec{\tilde{p}}_\psi}{\tilde{p}^0_\psi} = \vec{v}_\psi,\qquad
 \left\{  \begin{aligned}
&\frac{\d \vec{\tilde{p}}_\psi}{\d \tilde{t}} = -\frac{\vec{\tilde{\nabla}} \tilde{m}^2_{\rm eff}}{2 \tilde{p}^0_{\psi}}, \qquad &\text{Scalar} \\
&\frac{\d \vec{\tilde{p}}_\psi}{\d \tilde{t}} = \pm \, g_V (\vec{\tilde{E}}_{A^\prime} + \vec{v}_\psi\times\vec{\tilde{B}}_{A^\prime} ), \qquad \qquad &\text{Vector}
\end{aligned}\right. 
.\label{eq:traj}
\ee
Utilizing a Monte Carlo simulation, we generate fermion events, distributing initial positions within the range $2 < r/r_g < 40/\alpha^2$, the latter significantly exceeding the cloud’s Bohr radius. The generalized acceptance-rejection method~\cite{arm} is applied to a distribution weighted by the production rate $\Gamma_{S\psi} \propto \phi_0^{3/2}$ for scalar fields and $\Gamma_{V\psi} \propto E_{A^\prime}^2$ for vector fields. Setting the initial time of each event to $\tilde{t}=0$ for simplicity, fermion production from a scalar cloud is concentrated in planes at $\varphi = \pm \pi/2$, whereas vector cloud production spans the entire wavefunction.

To evolve the trajectory of each event using Eq.~(\ref{eq:traj}), we need to specify the initial dimensionless momentum on the order of $\sim \sqrt{\mu/(g_{S/V}\Psi_0)}$ and the fermion’s bare mass, $\tilde{m}_\psi$. In a scalar cloud, fermions are generated when $\tilde{m}_{\rm eff}$ is close to zero, with their initial momentum isotropically distributed. Conversely, in a vector cloud, the initial fermion momentum either aligns or anti-aligns with the electric fields. Both the initial momentum and $\tilde{m}_\psi$ are significantly less than $1$ in the parameter space relevant to this study. Testing for trajectory convergence demonstrated that outcomes are independent of the initial momentum values, as the bosonic cloud quickly adjusts $|\vec{\tilde{p}}_{\psi}|$ to $\mathcal{O}(1)$ values. Consequently, we choose a sufficiently small initial momentum and set both $\tilde{m}_{\rm eff}$ and $\tilde{m}_\psi$ to zero for the initial conditions.

\begin{figure}[h]
    \centering
\includegraphics[width=0.95\textwidth]{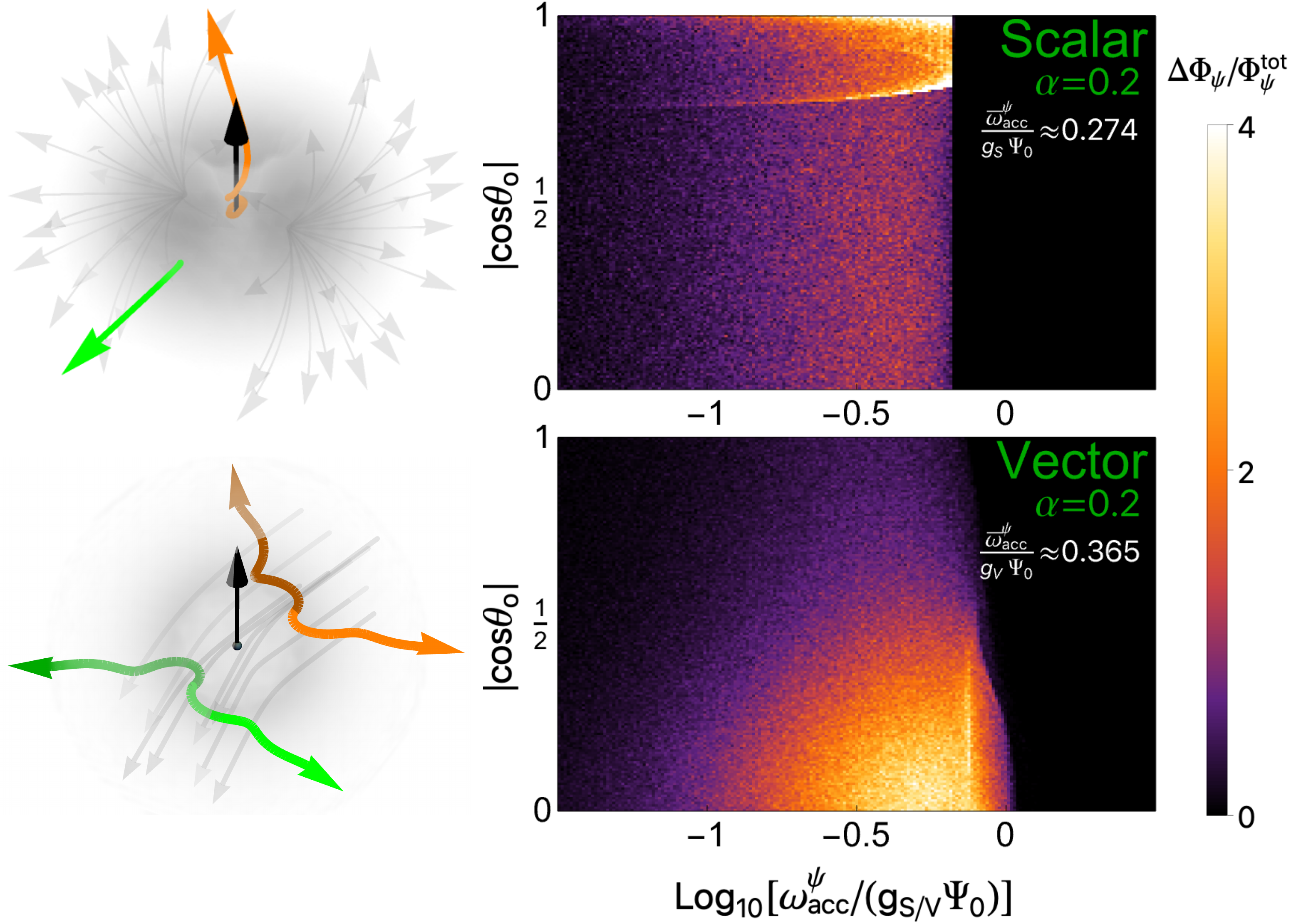}
    \caption{Top left: Examples of fermion trajectories originating simultaneously from distances of $10\,r_g$ (orange) and $100\,r_g$ (green) relative to a BH surrounded by a scalar cloud, with $\alpha$ set at $0.2$. The black arrow indicates the direction of the BH's spin, and the cloud's energy density alongside the initial scalar force lines are depicted in gray.
    Top right: The observed fermion energy spectrum $\Delta \Phi_\psi \equiv \d\Phi_\psi/(\d\cos\theta_o\,\d{\rm log}_{10}\omega_{\rm acc}^\psi)$ across various inclination angles $\theta_o$, with $\Phi_\psi^{\rm tot}$ representing the total emitted flux. The average energy of emitted fermions, $\bar{\omega}_{\rm acc}^\psi$, is on the order of $g_{S}\Psi_0$.
    Bottom: Analogous to the top panels but for a vector cloud. Here, trajectories in green and orange represent fermion paths from $30\,r_g$, with green tracing the equatorial plane and orange at $\theta = \pi/4$. Light and dark hues distinguish between fermions and anti-fermions, respectively.}
    \label{fig:nutrajflux}
\end{figure}

The trajectory analysis concludes at $\tilde{t}=400/\alpha^2$, where both the momentum and the outgoing direction $(\theta, \varphi)$ converge to nearly constant values. Assuming the observation duration exceeds the oscillation period $2\pi/\mu$, the impact of the azimuthal angle $\varphi$ on the event can be disregarded. For a vector cloud, fermion and anti-fermion fluxes are identical, as the trajectory of an anti-fermion mirrors that of a fermion originating from the same point but beginning a half-period $\pi/\mu$ later, aligning with a phase shift in the boson wavefunction. The final momentum and polar angle are documented as $\omega_{\rm acc}^\psi/(g_{S/V}\Psi_0) \equiv \tilde{p}^0_{\psi}$ and $\cos\theta_o \equiv \tilde{p}^z_\psi/\tilde{p}^0_{\psi}$, respectively.

\begin{figure}[h]
    \centering
    \includegraphics[width=0.95\textwidth]{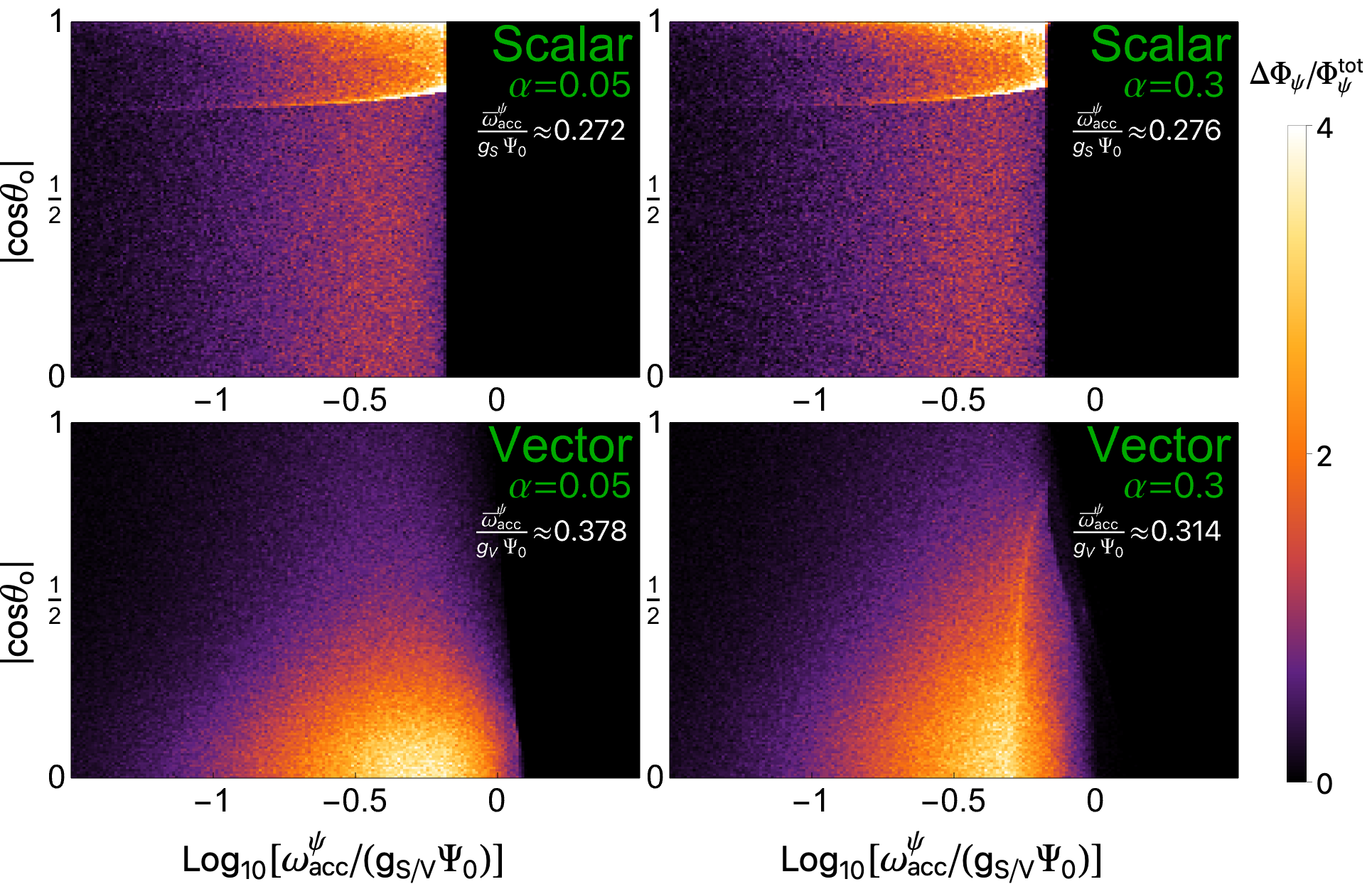}
    \caption{Same as Fig.~\ref{fig:nutrajflux} with $\alpha=0.05$ and $0.3$.}
    \label{SMfig:nutrajflux}
\end{figure}

An example depicted in the top left panel of Fig.~\ref{fig:nutrajflux} illustrates trajectories emanating from $10\,r_g$ (orange) and $100\,r_g$ (green) within a scalar cloud of $\alpha=0.2$. Additionally, we depict the force lines at the production time and scalar density in gray. Notably, the trajectory originating from the outer region exits the cloud directly in the radial direction, while the trajectory closer to the BH first curls around the BH spin axis (black arrow). This occurs because fermions produced at larger radii predominantly experience acceleration in the radial direction due to the first term on the right-hand side of Eq.~(\ref{app:terms}), resulting in an almost isotropic flux distribution for $|\cos\theta_o|<3/4$. Conversely, in the cloud’s inner region, the second term of Eq.~(\ref{app:terms}) becomes significant, confining trajectories to areas with small inclination angles ($|\cos\theta_o|>3/4$) and leading to a noticeable increase in fermion flux. The top right panel illustrates the outgoing spectrum as a function of the observer’s inclination angle, $\theta_o$, highlighting a jet-like structure of fermions with higher flux observed from nearly face-on angles ($|\cos\theta_o| > 3/4$), attributed to fermions produced closer to the cloud’s center being funneled along the polar axis. Our simulations indicate an average energy for the fermion fluxes of $\bar{\omega}^\psi_{\rm acc}/(g_S \Psi_0) \approx 0.274$. In Fig.~\ref{SMfig:nutrajflux}, we compare results for $\alpha = 0.05$ and $0.3$. Notably, both the spectrum and average momentum demonstrate minimal dependence on $\alpha$.

For the vector cloud, we illustrate examples of both fermion (light) and anti-fermion (dark) trajectories in the bottom left panel of Fig.~\ref{fig:nutrajflux}, with the outgoing spectrum showcased in the bottom right panel. The distribution of the spectrum across inclination angles $\theta_o$ reveals distinct behaviors between scalar and vector clouds. In vector clouds, acceleration is primarily influenced by electric fields, which are predominantly perpendicular, resulting in a greater flux directed towards observers in an edge-on position, where $\cos \theta_o \approx 0$. Although magnetic fields may deflect trajectories towards higher latitude regions, their influence is reduced by a factor of $\alpha$ compared to that of electric fields. Nevertheless, fluxes towards face-on observers ($|\cos\theta_o| \approx 1$) remain substantial due to these deflections. The $\alpha$-dependence is again minor, as demonstrated in Fig.~\ref{SMfig:nutrajflux}.

Through this analysis, the resultant average energy of the fermion fluxes, $\bar{\omega}^\psi_{\rm acc}$, is given by
\be \bar{\omega}^\psi_{\rm acc}  \approx \left\{\begin{array}{ll}
0.27\,g_{S}\,\Psi_0,\qquad \qquad &\text{Scalar} \\
0.35\,g_V\,\Psi_0,\qquad \qquad &\text{Vector}
\end{array}\right.
,\label{omeganuacc} \ee
which is generally independent of the parameter $\alpha$.

The differential fermion fluxes observable by a distant observer can be estimated as follows:
\be \label{eq:flux}  \begin{split}
 & \frac{\d \Phi_\psi}{\d \omega^\psi_{\rm acc}} (\bar{\omega}^\psi_{\rm acc})  \approx   \frac{2\,N_\psi \int \Gamma_{S\psi/V\psi}\, \d^3 \vec{x}}{4\pi d^2\,\bar{\omega}^\psi_{\rm acc}}\\
\approx &     {\footnotesize \left\{  \begin{aligned}
   &  1.2\times 10^{-8}\, {\rm cm^{-2}s^{-1}GeV^{-1}} \, \left(\frac{\Psi_0/\text{GeV}}{4.8\times10^{7}}\right)^{1/2}
 \,   \left(\frac{N_\psi}{3}\right) \, \left(\frac{10^{-12}}{\mu/\text{eV}}\right)^{1/2} \, \left(\frac{0.3}{\alpha}\right)^3 \, \left(\frac{g_{S}}{10^{-8}}\right)^{1/2} \,\left(\frac{5}{d/\text{kpc}}\right)^2,\\
& 1.3\times10^3 \ \ \,  {\rm cm^{-2}s^{-1}GeV^{-1}} \, \left(\frac{\Psi_0/\text{GeV}}{5.7\times10^{14}}\right)
 \ \ \ \left(\frac{N_\psi}{1}\right) \, \left(\frac{10^{-12}}{\mu/\text{eV}}\right) \ \ \ \ \left(\frac{0.3}{\alpha}\right)^3\, \left(\frac{g_{V}}{10^{-12}}\right)\ \ \  \,  \left(\frac{5}{d/\text{kpc}}\right)^2,
     \end{aligned} \right.}
\end{split}\ee
for scalar and vector clouds, respectively, where the boson mass and $\alpha$ values correspond to a BH of $40\,M_\odot$. Here, the factor of $2$ accounts for the production of fermion pairs, $N_\psi$ represents the number of fermion mass/flavor eigenstates being produced and accelerated, and $d$ is the distance between the BH and the observer. {The numerical values are evaluated using the expression in the first line, with benchmark parameters for both scalar and vector cases. The resulting spectrum is centered around $\bar{\omega}^\psi_{\rm acc}$, with a width of the same order, as shown in Figs.~\ref{fig:nutrajflux},~\ref{SMfig:nutrajflux}. Additionally, while Eq.~(\ref{eq:flux}) assumes isotropic emission, the observed flux can vary with the observer’s inclination angle, potentially fluctuating by an order of magnitude, as indicated in Figs.~\ref{fig:nutrajflux},~\ref{SMfig:nutrajflux}. The large discrepancy between the scalar and vector cases arises primarily from the field values in these benchmark parameters, which are chosen to satisfy the potential saturation conditions discussed in the following sections.}

Here, we adopt $N_\psi = 3$ for scalar fields and $N_\psi = 1$ for vector fields as our benchmark values to align with the neutrino analysis in subsequent sections. A deviation from these values does not significantly affect the qualitative results. Given the large galactic scale distance, approximately $d \sim$\,kpc, an approximately uniform distribution among the various flavor eigenstates can occur due to mixing.

\subsection{Saturating Cloud States}\label{sec:SCS}

The growth of a boson cloud may continue until the rate of energy loss from fermion emissions aligns with the rate of energy gain through BH superradiance. As discussed in Sec.~\ref{sec:FPSC}, the energy loss rate, scaled as $\bar{\omega}^\psi_{\rm acc} \Gamma_{S\psi/V\psi}$, results in $\Psi_0^{5/2}$ for scalar fields and $\Psi_0^{3}$ for vector fields. In contrast, the superradiant energy gain rate, given by $\Gamma_{\rm SR} M_{\rm cloud} \propto \Psi_0^2$, follows a lower power relation. Balancing these two processes determines a critical field value, $\Psi_0^c$, which can be dynamically reached from an initially exponential growth phase. During the saturation phase, the boson cloud functions to convert BH rotational energy into a steady stream of fermion emissions.

\begin{figure}[h]
    \centering
    \includegraphics[width=0.9\textwidth]{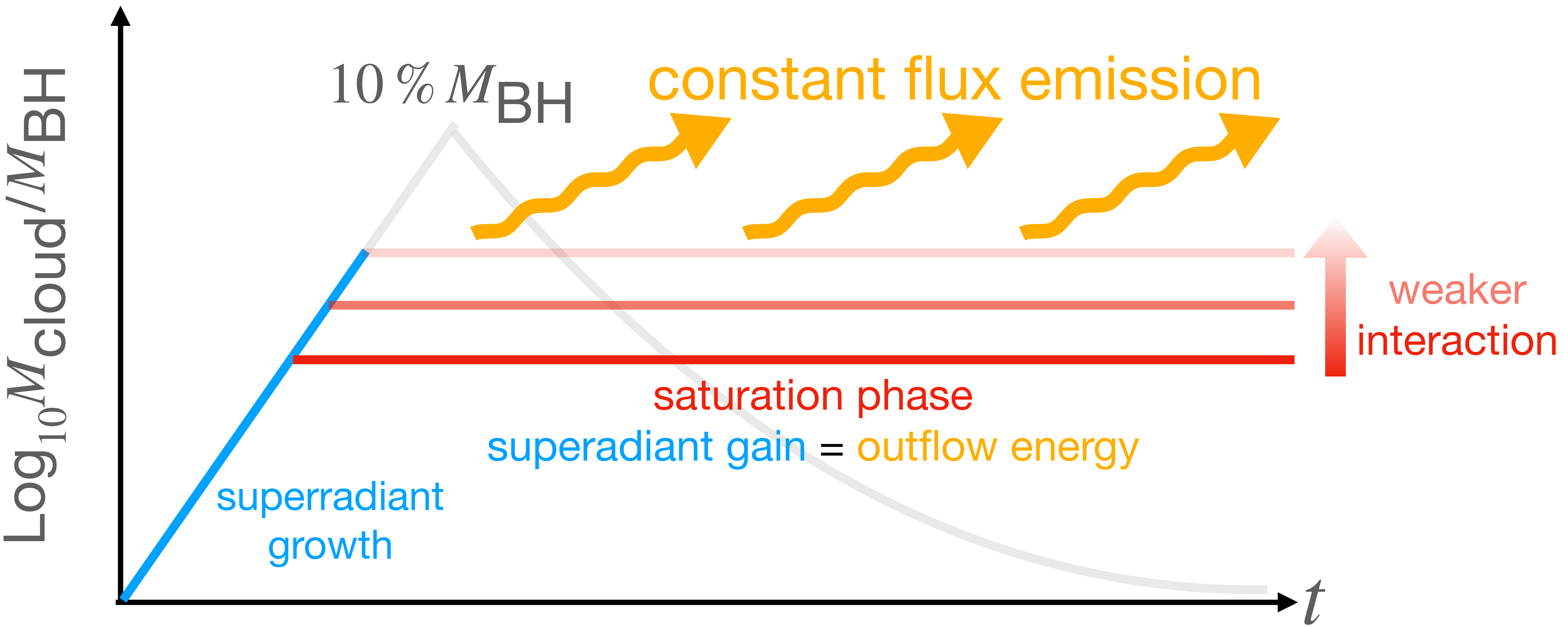}
    \caption{Illustration of an initially exponentially growing superradiant cloud transitioning into a saturation phase, depicted using the total cloud mass, $M_{\rm cloud}$, as a function of time. The blue portion represents the superradiant growth phase, while the red lines indicate saturation phases where flux emission balances the superradiant energy gain from the BH. A lower coupling constant, $g_{S/V}$, leads to a higher critical field value, $\Psi_0^c$, as detailed in Eq.~(\ref{A0c}).
    }
    \label{fig:Isaturation}
\end{figure}

This balance relation is expressed as:
\be \Gamma_{\rm SR}\, M_{\rm cloud} = 2\,N_\psi\, \bar{\omega}^\psi_{\rm acc} \int \Gamma_{S\psi/V\psi} \, \d^3 \vec{x}.\label{balcon}\ee
Here, the left side represents the superradiant energy gain rate, while the right side quantifies the energy loss rate due to fermion emission. Solving this equation yields the critical field $\Psi_0^c$:
\be
 {\Psi}_0^c \approx
\left\{  \begin{aligned}
   & 4.8\times10^{7}\, {\rm GeV} \, \ \left(\frac{3}{N_\psi}\right)^2  \left(\frac{\mu/\text{eV}}{10^{-12}}\right) \,\left(\frac{\alpha}{0.3}\right)^{16} \,  \left(\frac{10^{-7}}{g_{S}}\right)^5 \, \left(\frac{a_J}{0.9}\right)^2,\qquad  &\text{Scalar} \\
& 5.7\times10^{14}\, {\rm GeV} \, \left(\frac{1}{N_\psi}\right)\  \left(\frac{\mu/\text{eV}}{10^{-12}}\right)\, \left(\frac{\alpha}{0.3}\right)^6 \ \left(\frac{10^{-12}}{g_{V}}\right)^3\,\left(\frac{a_J}{0.9}\right), \qquad &\text{Vector}
      \end{aligned} \right. .
      \label{A0c}
      \ee
The benchmark parameters used here are in accordance with those from Eq.~(\ref{eq:flux}). Incorporating these parameters and the saturation field value from Eq.~(\ref{A0c}) into Eq.~(\ref{omeganuacc}) and Eq.~(\ref{eq:flux}), we predict a steady fermion emission characterized by a typical energy scale in the TeV range for the vector case. The resultant fluxes are estimated to be around $10^3\,\text{cm}^{-2}\text{s}^{-1}\text{GeV}^{-1}$, markedly surpassing the TeV-scale atmospheric neutrino flux~\cite{Vitagliano:2019yzm}. {The discrepancy between the scalar and vector cases stems from the chosen benchmark parameters. However, even if the parameters are adjusted to equalize $\Psi_0^c$, the vector field can still dominate the energy emission rate due to the superradiant rate ratio $\sim 96/\alpha^2$ in the saturated limit.}

An illustrative plot of this process is shown in Fig.~\ref{fig:Isaturation}, depicting the evolution of the total cloud mass over time. Initially, the cloud grows exponentially due to superradiance, then transitions to a saturation phase where steady fermion emissions balance the energy influx from BH superradiance. Notably, during the saturation phase, both the fermion fluxes and the average energy, $\bar{\omega}^\psi_{\rm acc}$, increase due to a lower coupling constant, which results in a higher critical field $\Psi_0^c$, as detailed in Eq.~(\ref{A0c}).

{The duration of the saturation phase can be estimated as $\sim M_{\rm BH}/(\Gamma_{\rm SR} M_{\rm cloud})$, where $\Gamma_{\rm SR} M_{\rm cloud}$ represents the linear energy extraction rate. For the scalar case, using the benchmark values in Eq.~(\ref{A0c}), the estimated timescale is several orders of magnitude longer than the age of the universe. For the vector case, the saturation phase lasts a few days due to the high benchmark field value. A shift in the benchmark parameter from $g_V = 10^{-12}$ to $10^{-11}$ significantly extends the saturation time to approximately a thousand years.}

Whether a saturation state dominated by fermion emissions can be achieved depends on several factors. A straightforward requirement is that the critical field value ${\Psi}_0^c$ should stay below $\Psi_0^{10\%}$, above which the saturation cannot be realized as the BHs' rotation already slows down. Another requirement comes from kinematics for parametric production—$g_{S}\Psi_0 \gg m_\psi$ for scalar fields and a more rigorous condition $\sqrt{g_{V}\Psi_0\mu} \gg m_\psi$ for vector fields.

Interactions with fermions lead to irreducible boson self-interactions that can quench exponential growth. For example, a Yukawa-type coupling generates a $\lambda \phi^4$ term through the Coleman-Weinberg mechanism, where $\lambda = g_S^4/(4\pi^2)$~\cite{Dev:2022bae}. When compared to a saturated axion state where $f_\phi = \mu/\sqrt{24 \lambda}$, the critical field value $\Psi_0^c$ should remain below $\Psi_0^\lambda \approx \alpha f_\phi/2 = \alpha \mu \pi/(2\sqrt{6} g_S^2)$. This requirement immediately rules out saturation under the benchmark parameters used in Eq.~(\ref{A0c}). Conversely, a vector field with a U(1) gauge interaction can induce only Euler-Heisenberg interactions, represented by $[(\vec{E}_{A^{\prime}}^2-\vec{B}_{A^{\prime}}^2)^2 + 7 (\vec{E}_{A^{\prime}} \cdot \vec{B}_{A^{\prime}})^2]  g_V^4/(360\pi^2 m_\psi^4)$~\cite{Heisenberg:1936nmg}. Due to the higher dimension operator, this results in a suppression factor of $\mu^4/m_\psi^4$ compared to the scalar case, making this type of self-interaction inadequate for quenching superradiance within the parameters considered in this study.

Finally, the Yukawa interaction between the scalar cloud and fermions can significantly increase the fermions’ effective mass, enabling decay processes that are kinetically forbidden in vacuum due to the low bare mass of the fermions. For instance, a standard model neutrino can decay into a charged pion and an electron or positron once its effective mass exceeds the mass of charged pions ($m_{\pi} \approx 140$\,MeV), with the decay rate quickly surpassing the boson oscillation frequency $\sim \mu$. Consequently, the neutrino remains non-relativistic during its decay, drawing energy from the BH approximately equal to $m_{\pi}$ rather than $g_{S}\Psi_0^c$. This scenario results in an energy loss rate proportional to $\Psi_0^{3/2}$, which is insufficient to balance the superradiant gain rate. Therefore, another condition emerges: $m_{\rm eff} \sim g_S \Psi_0^c \ll m_{\rm dec}$, where $m_{\rm dec}$ is the decay threshold. This issue does not occur with vector fields.

\section{Spin Measurements for Neutrino-Boson Interaction}\label{sec:SMNBI}
We now turn our attention to constraints on boson-fermion interactions. Within the Standard Model of particle physics, the coupling of most fermions to hidden ultralight bosons has been stringently constrained by experiments testing the fifth force, the equivalence principle, atomic clocks, and astrometry~\cite{Graham:2015ifn,Berge:2017ovy,Hees:2018fpg,Pierce:2018xmy,Kennedy:2020bac,LIGOScientific:2021ffg,Shaw:2021gnp,PPTA:2021uzb,Ge:2021lur,Yu:2023iog,NANOGrav:2023hvm}. These constraints typically place $g_{S/V}$ well below $10^{-20}$, making the critical field value $\Psi_0^c$ from Eq.~(\ref{A0c}) significantly exceed $\Psi_0^{10\%}$, thereby preventing the saturation state from being reached. However, the mechanism of fermion production and acceleration by a bosonic cloud also extends to hidden sectors, where the coupling strength is often unconstrained.

A particularly notable case is that of the Standard Model neutrino, whose interactions with ultralight bosons are exceptionally challenging to detect due to difficulties in production, manipulation, and detection. We consider both scalar and vector interactions with neutrinos. The scalar field, often postulated as the Majoron, is motivated by mechanisms of neutrino mass generation~\cite{Gelmini:1980re,Chikashige:1980ui,Aulakh:1982yn}, while the vector field stems from grand unification theories~\cite{Georgi:1974sy,Pati:1974yy,Mohapatra:1974hk,Fritzsch:1974nn,Georgi:1974my}. The interactions between these fields and the Standard Model neutrino, $\nu_L$, in two-component spinor notation, are described as follows:~\cite{Berryman:2022hds}
\be \text{Scalar}:\ g_{S\nu}\, \phi\, \nu_L \nu_L; \qquad \text{Vector}:\ g_{V\nu} A^{\prime\mu} {\nu}^\dagger_L \bar{\sigma}^\mu \nu_L.\label{eq:gAnu}
\ee
We assume a universal coupling strength, $g_{S\nu}$ or $g_{V\nu}$, across all mass or flavor eigenstates. {Note that the Majoron $\phi$ is typically assigned a lepton number of $L = -2$, while each $\nu_L$ carries $L = +1$.}

Model-independent constraints on these couplings, particularly for ultralight bosons, primarily arise from neutrino self-interactions mediated by these bosons~\cite{Berryman:2022hds}. These constraints are largely unaffected by the boson’s mass $\mu$ when it is below the eV scale, and have been established through various observations including supernova (SN) 1987A~\cite{Kolb:1987qy,Kachelriess:2000qc,Farzan:2002wx}, double beta decay~\cite{KamLAND-Zen:2012uen,Brune:2018sab}, big bang nucleosynthesis~\cite{Huang:2017egl}, and the cosmic microwave background (CMB)~\cite{Forastieri:2019cuf,Li:2023puz}. Specifically, upper limits of $g_{S\nu} < 3 \times 10^{-7}$ have been inferred from SN 1987A~\cite{Kachelriess:2000qc,Farzan:2002wx}, and $g_{V\nu} < 7 \times 10^{-7}$ from the CMB~\cite{Forastieri:2019cuf}. Above the eV threshold, the variability of constraints in relation to the boson’s mass $\mu$ significantly increases~\cite{Escudero:2019gvw,Fiorillo:2022cdq,Sandner:2023ptm,Fiorillo:2023cas,Fiorillo:2023ytr}.

There are scenarios where these couplings are subject to more stringent constraints: If bosons constitute a significant portion of dark matter, they could potentially alter neutrino oscillation patterns~\cite{Reynoso:2016hjr,Berlin:2016woy,Krnjaic:2017zlz,Brdar:2017kbt,Davoudiasl:2018hjw,Liao:2018byh,Capozzi:2018bps,Huang:2018cwo,Farzan:2019yvo,Cline:2019seo,Dev:2020kgz,Losada:2021bxx,Huang:2021kam,Chun:2021ief,Dev:2022bae,Huang:2022wmz,Losada:2022uvr,Brzeminski:2022rkf,Alonso-Alvarez:2023tii,ChoeJo:2023ffp,Losada:2023zap}. Interactions between bosons and other Standard Model fermions also impose additional constraints~\cite{Graham:2015ifn,Pierce:2018xmy,LIGOScientific:2021ffg,Shaw:2021gnp,PPTA:2021uzb,Ge:2021lur,Yu:2023iog}. Furthermore, when the fermion current is not conserved under the vector boson’s gauge group, physical processes are predominantly influenced by the vector’s longitudinal mode, leading to divergent constraints for an ultralight boson mass $\mu$~\cite{Laha:2013xua,Bakhti:2017jhm,Escudero:2019gzq,Bahraminasr:2020ssz,Dror:2020fbh,Ekhterachian:2021rkx,Singh:2023nek}. However, superradiant bosons need not make up the majority of dark matter, and through sophisticated model building, it is feasible to circumvent couplings to other Standard Model fermions and mitigate issues caused by the longitudinal mode. Therefore, our analysis will focus on model-independent constraints stemming from neutrino self-interactions.

Neutrino pairs can spontaneously emerge within both scalar and vector bosonic clouds, facilitated by light neutrino masses—the sum of the three mass eigenstates is less than $0.12$\,eV~\cite{Planck:2018vyg} and the lightest is unbounded from below~\cite{GAMBITCosmologyWorkgroup:2020rmf}—coupled with the substantial field values of the bosons. The kinematic threshold for parametric production from scalar clouds, $g_{S\nu}\Psi_0 \gg m_\nu$, can be easily satisfied for all three neutrino mass eigenstates. For vector fields, the more stringent condition $\sqrt{g_{V\nu}\Psi_0\mu} \gg m_\psi$ is achievable for the lightest neutrino mass eigenstate in superradiant clouds around stellar-mass BHs. Consequently, we adopt $N_\nu = 3$ for scalar fields and $N_\nu = 1$ for vector fields as our benchmark values for $N_\psi$ in Eqs.~(\ref{eq:flux},\ref{A0c}), which have minor impacts on the quantitative results. At a galactic scale distance, $d \sim$\,kpc, the flux distribution among the three neutrino flavors tends to be approximately uniform due to neutrino oscillation effects.

\begin{figure}[thb]
    \centering
    \includegraphics[width=0.9\textwidth]{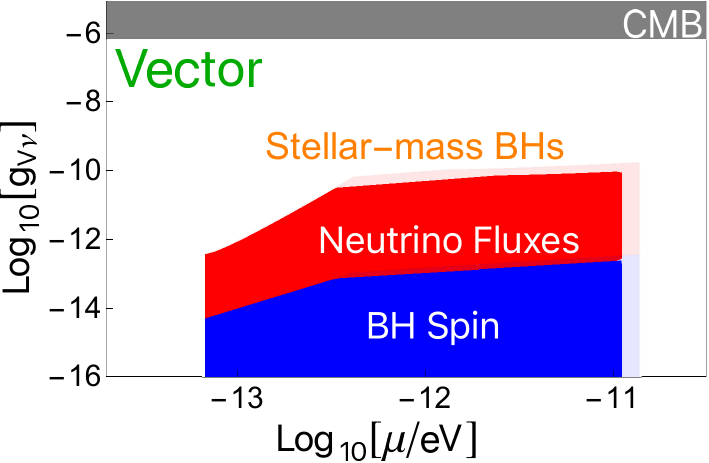}
    \caption{Prospects for constraining neutrino-vector couplings, $g_{V\nu}$. We consider a stellar-mass BH population with masses ranging from $3\,M_\odot$ to $100\,M_\odot$. The blue region marks the exclusion zone derived from BH spin measurements, with darker shades for a spin parameter $a_J= 0.8$ and lighter shades extending to $a_J= 0.9$, highlighting the impact of spin measurement uncertainties. The red region denotes areas where observable neutrino fluxes are expected from a target BH situated at a distance of $5$\,kpc. For reference, previous exclusions based on observations from the CMB~\cite{Forastieri:2019cuf} are shown in gray.
    }
    \label{fig:constraints}
\end{figure}

To ensure uninterrupted superradiant energy transfer, the critical field value $\Psi_0^c$ must remain below $\Psi_0^{10\%}$ to mitigate strong backreaction effects on BH rotation. This requirement defines an exclusion zone in parameter space for fast-rotating BHs that do not meet this criterion, particularly when the superradiant timescale is notably short compared to the BH’s evolutionary time. The focus is on rapidly rotating BHs, with confirmed masses ranging from $3\,M_\odot$ to approximately $100\,M_\odot$, evidenced by recent observations including X-ray binaries with significant separations~\cite{1997xrb..book.....L}, microlensing~\cite{Mao:2001xs,Bennett:2001vh}, gravitational wave observations~\cite{Barack:2018yly}, and spectroscopic data such as iron lines that further inform on BH spins~\cite{Brenneman:2006hw}. Moreover, the Event Horizon Telescope, with its unprecedented angular resolution, has captured horizon-scale images of two nearby supermassive BHs: M87$^\star$ ($6.5\times 10^9\,M_\odot$)~\cite{EventHorizonTelescope:2019dse} and Sgr A$^\star$ ($4.3\times 10^6\,M_\odot$)~\cite{EventHorizonTelescope:2022wkp}, supporting high spins for these BHs. In our analysis, we consider dimensionless spins $a_J$ ranging from $0.8$ to $0.9$, with corresponding maximum $\alpha$ values of $0.25$ and $0.31$ for these spins, respectively.

We consider the exclusion criterion where $\Psi_0^c > \Psi_0^{10\%}$, which allows high-spin BHs to rule out the corresponding superradiant parameter space. In Fig.~\ref{fig:constraints}, we illustrate the spin measurement constraints for the vector-neutrino coupling $g_{V\nu}$ using the population of stellar-mass BHs. The dark blue region denotes constraints for a dimensionless spin parameter $a_J = 0.8$, while the light blue represents the expanded exclusion zone for $a_J = 0.9$. The analysis begins with a lower limit for $\alpha$ at $0.05$. The boson mass, defined as $\mu \equiv \alpha/(G_N M_{\rm BH})$, is divided into two regions. In the low-mass segment, we set $M_{\rm BH} = 100\,M_\odot$ and vary $\alpha$ from $0.05$ up to its upper limits of $0.25/0.31$. In the high-mass section, $\alpha$ is fixed at $0.25/0.31$, and $M_{\rm BH}$ is adjusted from $100\,M_\odot$ down to $3\,M_\odot$. For SMBHs, boson masses below $10^{-17}$\,eV face challenges in meeting the kinetic conditions for relativistic Schwinger pair production, making them difficult to consider in this context. This quantitative exclusion also applies to dark fermions whose mass meets the kinematic threshold.

The dynamics of the scalar-neutrino interaction, as discussed in Sec.~\ref{sec:SCS}, involve complex phenomena including quartic self-interactions induced by neutrino loops. This interaction introduces a threshold $\Psi_0^\lambda = \alpha \mu \pi/(2\sqrt{6} g_{S\nu}^2)$, which is significantly lower than $\Psi_0^c$ when $g_{S\nu} \ll 10^{-7}$. As a result, a saturation state driven by scalar quartic self-interactions is expected, with predominant fluxes from scalar emissions rather than neutrino emissions. Therefore, spin measurements can only exclude regions where $\Psi_0^\lambda > \Psi_0^{10\%}$, which translates to weak constraints on $g_{S\nu}$, typically $g_{S\nu} \ll 10^{-18}$ for boson masses around $10^{-12}$\,eV, and even weaker for scalar masses associated with SMBHs.

\section{Neutrino and Boosted Dark Matter Fluxes from Superradiant Clouds}\label{sec:NBDMF}

For couplings beyond the limits set by spin measurements, a saturated vector cloud consistently emitting neutrino fluxes is expected. Once the critical field value, $\Psi_0^c$, is determined from Eq.~(\ref{A0c}), the neutrino flux spectrum for a specific BH can be calculated using Eq.~(\ref{eq:flux}), factoring in the BH’s distance from Earth $d$, and the observer’s inclination angle $\theta_o$. Considering a benchmark scenario with parameters $\mu = 10^{-12}$\,eV, $M_{\rm BH} = 40\,M_{\odot}$, $d=5$\,kpc, $g_{V\nu} = 10^{-12}$, and $a_J = 0.9$, we find $\Psi_0^c\approx 5.7\times 10^{14}$\,GeV. This leads to TeV-scale neutrino fluxes around $10^3$\,cm$^{-2}$s$^{-1}$GeV$^{-1}$, substantially exceeding the atmospheric neutrino flux at TeV energies~\cite{Vitagliano:2019yzm}. The neutrino fluxes and their average energy increase during the saturation phase as $g_{V\nu}$ decreases, attributable to a lower $g_{V\nu}$ yielding a higher critical amplitude $\Psi_0^c$.

With the cessation of exponential BH spin energy extraction by the cloud, a high-spin BH can coexist with a saturated cloud. Consequently, we maintain $a_J$ values ranging from $0.8$ to $0.9$, while $\mu$ is categorized as per the spin measurement analysis. Given the wide distribution of BH candidates throughout the Milky Way, we set a standard distance of  $d = 5$\,kpc for analyzing the fluxes from a target BH. In Fig.~\ref{fig:constraints}, the red region denotes where neutrino fluxes from a saturated cloud are observable. The upper boundary is determined by requiring that neutrino fluxes at $\bar{\omega}^\nu_{\rm acc}$ exceed $1\%$ of the diffusive neutrino background~\cite{Vitagliano:2019yzm}, aligned with the current IceCube angular resolution of $<10^\circ$~\cite{IceCube:2018ndw,IceCube:2023ame}.

Neutrino production from a bosonic cloud is also applicable to other coupled fermions in the hidden sector. Moreover, dark matter particles within the vicinity of the cloud can be directly accelerated to higher energies. Thus, the superradiant cloud can act as a source for boosted dark matter, presenting opportunities for detection through direct detection experiments and neutrino detectors~\cite{DEramo:2010keq,Huang:2013xfa,Agashe:2014yua,Kopp:2015bfa,Bhattacharya:2016tma,Kamada:2017gfc,Kachulis:2017nci,Chatterjee:2018mej,Kamada:2018hte,McKeen:2018pbb,Arguelles:2019xgp,Kamada:2019wjo,Berger:2019ttc,Abi:2020evt}. The specific incoming directions of these particles could lead to daily modulation of signals due to their interaction with Earth materials~\cite{Ge:2020yuf,Fornal:2020npv,Chen:2021ifo,Xia:2021vbz,PandaX-II:2021kai,Qiao:2023pbw} or when captured by directional detectors~\cite{Vahsen:2021gnb}.

In the case of scalar clouds, while scalar emissions are dominant, steady subleading fermion fluxes can also be expected. Detecting both spectra could help elucidate the origin of scalar self-interactions.

\section{Discussion}\label{sec:Discussion}

Strong fields serve as a fertile ground for exploring various fascinating phenomena, notably particle production. These phenomena have been studied within the early universe’s preheating phase~\cite{Greene:1998nh,Greene:2000ew} and the domain of strong field quantum electrodynamics~\cite{Schwinger:1951nm}. Rotating BHs offer an exceptional platform for such explorations, especially as superradiant clouds around them can attain field values nearing the Planck scale.

This study examines the interactions between ultralight bosonic fields and fermions facilitated by BH superradiance. Dense superradiant bosonic clouds are shown to efficiently generate and accelerate fermions, encompassing both scalar and vector types. We specifically delve into neutrino-boson interactions, approached from two angles: BH spin measurements can exclude interactions with very weak couplings, while observations of neutrino fluxes from high-spin BHs can help define the boundaries of strong interactions.

We demonstrate that neutrino fluxes emitted from a nearby stellar-mass BH, surrounded by a vector cloud, can significantly surpass the diffusive background~\cite{IceCube:2015mgt,Super-Kamiokande:2015qek,ANTARES:2017srd,Vitagliano:2019yzm,Baikal-GVD:2022fis}. Additionally, interactions between bosons and hidden-sector fermions could result in the emission of boosted dark matter fluxes. This opens a new frontier in multi-messenger astronomy: electromagnetic and gravitational wave observations identify BHs and ascertain their masses, spins, and inclination angles, while neutrino~\cite{IceCube:2018ndw,IceCube:2023ame,Illuminati:2023app,Baikal-GVD:2023pah,KM3Net:2016zxf,P-ONE:2020ljt,IceCube-Gen2:2020qha,TRIDENT,IceCube:2013dkx,KM3NeT:2018wnd,Fiorillo:2022ijt} and dark matter detectors~\cite{DEramo:2010keq,Huang:2013xfa,Agashe:2014yua,Kopp:2015bfa,Bhattacharya:2016tma,Kamada:2017gfc,Kachulis:2017nci,Chatterjee:2018mej,Kamada:2018hte,McKeen:2018pbb,Arguelles:2019xgp,Kamada:2019wjo,Berger:2019ttc,Abi:2020evt,Ge:2020yuf,Fornal:2020npv,Chen:2021ifo,Xia:2021vbz,PandaX-II:2021kai,Qiao:2023pbw} focus on these BHs to monitor the resulting flux emissions.

The current generation of neutrino detectors, such as IceCube~\cite{IceCube:2018ndw,IceCube:2023ame}, ANTARES~\cite{Illuminati:2023app}, and Baikal-GVD~\cite{Baikal-GVD:2023pah}, has already shown significant sensitivity in searching for point-like sources. Future detectors, including KM3NeT~\cite{KM3Net:2016zxf}, P-ONE~\cite{P-ONE:2020ljt}, IceCube-Gen2~\cite{IceCube-Gen2:2020qha}, and TRIDENT~\cite{TRIDENT}, are expected to further enhance angular resolution and offer more comprehensive sky coverage. Their performance is notably improved when integrated into a global network, boosting our capability to detect and analyze these intriguing phenomena~\cite{Schumacher:2021hhm}.

\section*{Acknowledgements} 
We are grateful to Markus Ahlers, Zhaoyu Bai, Mauricio Bustamante, Shao-Feng Ge, Minyuan Jiang, Qinrui Liu, Yuxin Liu, Luca Visinelli, Xin Wang, Donglian Xu and Xunjie Xu for useful discussions. This work is supported by VILLUM FONDEN (grant no. 37766), by the Danish Research Foundation, and under the European Union’s H2020 ERC Advanced Grant “Black holes: gravitational engines of discovery” grant agreement no. Gravitas–101052587. 
X.X. is supported by  Deutsche Forschungsgemeinschaft under Germany’s Excellence Strategy EXC2121 “Quantum Universe” - 390833306. Views and opinions expressed are however those of the author only and do not necessarily reflect those of the European Union or the European Research Council. Neither the European Union nor the granting authority can be held responsible for them. This work was supported by FCT (Fundação para a Ciência e Tecnologia I.P, Portugal) under project No.~2022.01324.PTDC. This project has received funding from the European Union's Horizon 2020 research and innovation programme under the Marie Sklodowska-Curie grant agreement No 101007855.


\end{document}